\pdfoutput=1
\documentclass{article}

\PassOptionsToPackage{numbers,compress}{natbib}
\usepackage[preprint]{neurips_2026}

\usepackage[utf8]{inputenc} % allow utf-8 input
\usepackage[T1]{fontenc}    % use 8-bit T1 fonts
\usepackage{hyperref}       % hyperlinks
\usepackage{url}            % simple URL typesetting
\usepackage{xurl}           % allow long URLs to break across lines
\usepackage{booktabs}       % professional-quality tables
\usepackage{amsfonts}       % blackboard math symbols
\usepackage{amsmath}
\usepackage{nicefrac}       % compact symbols for 1/2, etc.
\usepackage{microtype}      % microtypography
\usepackage{xcolor}         % colors
\usepackage{algorithm}
\usepackage{algpseudocode}
\usepackage{graphicx}

\title{Estimating Spectrum-Free Information Bounds for Generative Surrogates Beyond Conditional Collapse}

\author{%
  Chaitanya Dumbre \\
  School of Computer Science and Engineering \\
  MIT World Peace University, Pune \\
  \texttt{chaidumb04@gmail.com} \\
  \And
  Sourabh Dube \\
  Department of Physics \\
  Indian Institute of Science Education and Research (IISER), Pune \\
  \texttt{sdube@iiserpune.ac.in} \\
}

\begin{document}

\maketitle
\let\oldthefootnote\thefootnote
\renewcommand{\thefootnote}{}%
\footnotetext{Code: \href{https://github.com/NotArsenic/Estimating-Spectrum-Free-Information-Bounds-for-Generative-Surrogates-Beyond-Conditional-Collapse}{GitHub}}%
\let\thefootnote\oldthefootnote

\begin{abstract}
Conditional generative surrogates in high-energy physics, such as models that predict reconstructed jets from partons, are often evaluated on $1\text{D}$ marginal divergences among other metrics. Taken alone these metrics can reward conditional collapse. On a CMS simulation benchmark, a sampler that ignores the conditioner beats a trained Mixture Density Network (MDN) on all four target marginals under both KL and Wasserstein-1, and cannot be stably separated from a Conditional Flow Matching (CFM) model on two of the four. We instead score the models with the Continuous Ranked Probability Skill Score (CRPSS) from forecast verification, taking the exact marginal as the reference. We show how global skill depends on the evaluation spectrum, and we also report it within bins of a conditioning variable, where that dependence largely goes away. We then introduce a data-driven $k$-Nearest-Neighbor ($k$-NN) extrapolation that estimates the best CRPS the conditioning information allows. Both models agree with this estimate on all four targets, which suggests that current architectures have reached the information available in the conditioner, at least as far as these per-target scores can tell.
\end{abstract}

\section{Introduction}
\label{sec:intro}
Conditional generative models are increasingly explored as fast replacements for expensive scientific simulators such as {\sc geant4} \citep{allison_recent_2016} and multi-leg Monte Carlo generators \citep{alwall_automated_2014}, within the broader framework of simulation-based inference \citep{cranmer_frontier_2020,brehmer_mining_2020,paganini_calogan_2018,adelmann_new_2022,vaselli_flashsim_2024,butter_how_2019,theatlascollaboration_deep_2024}. The same conditioning input $c$ can lead to many different outcomes $y$, so such a model has to learn a conditional distribution $p(y \mid c)$ \citep{tarantola_inverse_2005}. Optimizing for mean squared error \citep{bishop_pattern_2016} instead collapses predictions to the conditional mean and compresses the generated marginal spectrum \citep{baattrup_pointwise_2026}. A common evaluation asks how well the marginal distribution of each output is reproduced, using histogram separation power, the Wasserstein-1 distance \citep{KR:58} or the Kullback--Leibler divergence \citep{kullback_information_1951}. Such comparisons appear throughout community benchmark suites \citep{kansal_jetnet_2023,krause_calochallenge_2025} and surrogate studies \citep{mikuni_score-based_2022,buhmann_caloclouds_2023,amram_denoising_2023}, usually alongside classifier two-sample tests \citep{krause_fast_2023,mikuni_score-based_2022,amram_denoising_2023} and Fréchet and kernel physics distances \citep{kansal_evaluating_2023,kansal_jetnet_2023}. Taken alone, a marginal comparison cannot show that a model uses $c$ at all, and a model that ignores $c$ is said to be conditionally collapsed.

We quantify this on the task of predicting reconstructed jet kinematics from parton kinematics in a CMS \citep{collaboration_cms_2008} QCD simulation dataset \citep{cms_collaboration_qcd_pt-15to7000_tunecp5_flat2018_13tev_pythia8runiisummer20ul16nanoaodv9-106x_mcrun2_asymptotic_v17-v1nanoaodsim_2024}. A condition-blind sampler, which reproduces the marginal by construction, beats a trained Mixture Density Network (MDN) \citep{bishop_mixture_1994} on 8 of 8 marginal-divergence comparisons across three data splits, by margins of $3.65\text{--}22.39\times$. A Conditional Flow Matching (CFM) surrogate \citep{lipman_flow_2023} matches the marginals more closely, but the sampler still beats it at every split on two of the four targets, and on the other two the winner changes from split to split.

This work builds on \citet{baattrup_pointwise_2026}, who show that pointwise metrics such as the root-mean-square error reward collapse onto the conditional mean in multimodal inverse problems. They propose evaluating such models with a per-event Continuous Ranked Probability Score (CRPS), a population-level test of spectrum fidelity and a coverage-based calibration check. Their setting is an inverse problem, but the argument holds for any model of $p(y \mid c)$ learned from simulated pairs, including a forward surrogate like ours. Marginal divergences fail in the opposite way, because they reward collapse onto the marginal. A condition-blind sampler passes a spectrum test by construction, and its prediction sets also have the right coverage when averaged over all events. In that protocol, only the per-event score and calibration checked within bins of phase space can tell it apart from a model that uses $c$. On its own, however, a CRPS value is hard to read. We give it a scale by taking the exact marginal as the zero of skill and a data-driven estimate of the best achievable score as the ceiling, and we show how the resulting skill depends on the evaluation spectrum.

This paper makes the following points.
\begin{itemize}
\item \textbf{The metric trap.} A condition-blind sampler beats a trained MDN on all four target marginals under both KL and $W_1$, by up to $22.4\times$, and cannot be stably separated from a CFM on two of four targets (Section~\ref{sec:metric_trap}).
\item \textbf{A reference from forecast verification.} We carry over the CRPS skill score with the exact marginal reference, which is the uncertainty term of the Hersbach decomposition \citep{hersbach_decomposition_2000}, and use it to fix the zero of the skill scale (Section~\ref{sec:CRPS_framework}).
\item \textbf{Spectrum dependence.} It is known that a climatological reference depends on how it is stratified \citep{murphy_skill_1988,hersbach_decomposition_2000}. We show what this does to global skill on our jet benchmark, and we report a differential version in bins of parton $p_{\mathrm{T}}$ alongside it (Sections~\ref{sec:CRPS_framework} and~\ref{sec:Target_dependent_saturation}).
\item \textbf{Data-driven limits.} A model-free $k$-NN extrapolation estimates the best achievable score, and both surrogates agree with it (Sections~\ref{sec:KNN} and~\ref{sec:Target_dependent_saturation}).
\item \textbf{Likelihood and CRPS differ.} Past the best epoch the likelihood degrades while the CRPS barely moves, so a flat CRPS alone does not rule out overfitting (Section~\ref{sec:Metric_decoupling}).
\end{itemize}

\section{Related Work}
\label{sec:related}
The weakness of marginal metrics is well known. In high-energy physics, \citet{kansal_evaluating_2023} point out that 1D binned projections lose sensitivity to correlations and conditional distributions, and they call for multivariate tests that include the conditioning features when a model is conditional. In image generation the same worry led to intra-class FID \citep{miyato_cgans_2018}, the Fréchet Joint Distance \citep{devries_evaluation_2019}, conditional FID \citep{soloveitchik_conditional_2021} and the Classification Accuracy Score \citep{ravuri_classification_2019}. Classifier two-sample tests \citep{lopez-paz_revisiting_2017} can detect collapse when the conditioning variable is given to the classifier, as in several calorimeter studies \citep{krause_fast_2023,krause_calochallenge_2025,amram_denoising_2023}. Comparing generated quantities against the conditioning variable exposes it directly \citep{mikuni_score-based_2022}, and the CaloChallenge also compares histograms of the total deposited energy divided by the incident energy, which probe the calorimeter response \citep{krause_calochallenge_2025}. Our momentum target is a ratio of the same kind, and the sampler still matches it, because a histogram pooled over all events tests the response averaged over $c$ and says little about how it changes with $c$. Fréchet and kernel physics distances computed on pooled samples do not see which event a prediction was conditioned on, although they can be extended to conditional generation \citep{kansal_evaluating_2023}. Simulation-based inference checks conditional calibration with rank and coverage tests \citep{talts_validating_2018,lemos_sampling-based_2023}.

Forecast verification has used proper scoring rules \citep{gneiting_strictly_2007} and skill scores against a climatological reference \citep{murphy_skill_1988,ferro_effect_2008} for decades, and most of our reference construction comes from that literature. \citet{murphy_skill_1988} separates single-valued references from multiple-valued ones built on subsamples of the data, and shows that forecasts generally look less skillful against the latter. \citet{hersbach_decomposition_2000} decomposes the CRPS into reliability, resolution and uncertainty, gives the sorted formula for the uncertainty, and notes that it falls when the climatology is stratified by location and season. Using the outcomes of similar past cases as a predictive ensemble is also old in weather forecasting, where it is known as an analog ensemble \citep{delle_monache_probabilistic_2013}. Our ceiling applies the same idea to the training set. The nearest training events of each test event serve as its ensemble, and extrapolating their score to zero neighbourhood radius estimates the best CRPS the conditioner allows, in the spirit of the nearest-neighbour bounds on the Bayes error of \citet{cover_nearest_1967}.

\citet{baattrup_pointwise_2026} recently proposed the CRPS as the per-event part of an evaluation protocol for inverse problems in particle physics, alongside spectrum fidelity and coverage-based calibration, and showed on dileptonic top-quark reconstruction that model rankings reverse between pointwise and distributional metrics. Our floor and ceiling give that per-event score a reference scale.

\section{Task and Data}
\label{app:physics_task}
The models are trained to predict the kinematic 4-vectors of reconstructed jets from the 4-vectors of the underlying partons. We use a public CMS QCD dataset \citep{cms_collaboration_qcd_pt-15to7000_tunecp5_flat2018_13tev_pythia8runiisummer20ul16nanoaodv9-106x_mcrun2_asymptotic_v17-v1nanoaodsim_2024}. Its events were generated with {\sc Pythia} 8 \citep{sjostrand_introduction_2015} using the CMS CP5 tune \citep{the_cms_collaboration_extraction_2020}, passed through the full {\sc geant4} \citep{allison_recent_2016} detector simulation and CMSSW reconstruction chain, and distributed as NanoAOD. The sample has a flat transverse momentum $\hat{p}_{\mathrm{T}}$ spectrum over a broad kinematic range. We keep reconstructed jets that satisfy $p_{\mathrm{T}}> 50$ GeV and with pseudorapidity ($\eta$) satisfying $\vert\eta\vert < 2.4$ that pass the jet identification requirement ($\texttt{jetId} \ge 2$). The jets are also required to satisfy pileup jet identification ($\texttt{puId} \ge 1$). On the parton side we take the two leading outgoing partons of the hard process and restrict their $p_{\mathrm{T}}$ to the fiducial window $[50, 1000)$ GeV. A parton and a jet form a pair when each is the other's nearest neighbour in $\Delta R$\footnote{$\Delta R = \sqrt{\Delta\eta^2+\Delta\phi^2}$, where $\phi$ is the azimuthal angle of the jet.} and $\Delta R < 0.2$. After the filters, about 10.8 million parton-jet pairs survive. We call each pair an event in what follows, and the data splits keep all pairs from one collision event together.
\subsection{The Inputs (The Conditioner $c$)}
Partons are close to massless at this kinematic scale ($p_{\mathrm{T}} \ge 50$ GeV), so the conditioning state is fixed by a spatial 3-vector and a discrete flavor. The models receive a 26-dimensional conditioning vector made of four blocks.
\begin{itemize}
  \item \textbf{Standardized kinematics (3 dims).} The parton's transverse momentum ($\log p_{\mathrm{T}}$), absolute pseudorapidity ($|\eta|$), and $\log(\cosh \eta)$, all $z$-score standardized.
  \item \textbf{Pileup (1 dim).} The number of true interactions (\texttt{nTrueInt}), $z$-score standardized. All results reported here use the recorded \texttt{nTrueInt} of each test event. When jets are generated from parton inputs alone, the value is instead sampled from the empirical pileup profile of the training set.
  \item \textbf{Flavor (6 dims).} One-hot vectors for the parton's identity ($d, u, s, c, b, g$).
  \item \textbf{Fourier-expanded azimuth (16 dims).} The periodic angle $\phi$ is expanded into 16 Fourier features, $\sin(k\phi)$ and $\cos(k\phi)$ for $k \in \{1, \dots, 8\}$.
\end{itemize}

\subsection{The Outputs (The Targets $y$)}
The models predict a 4-dimensional target $y$ that describes the residual between the parton and the reconstructed jet. It is built from the following transformations.
\begin{align}
  \text{Momentum response:} \quad            & y_0 = \log(p_{\mathrm{T}}^{\mathrm{jet}} / p_{\mathrm{T}}^{\mathrm{parton}}) \label{eq:target_pt}                             \\
  \text{Parity-folded pseudorapidity:} \quad & y_1 = \operatorname{sign}(\eta^{\mathrm{parton}}) \times (\eta^{\mathrm{jet}} - \eta^{\mathrm{parton}}) \label{eq:target_eta} \\
  \text{Azimuthal residual:} \quad           & y_2 = \operatorname{wrap}(\phi^{\mathrm{jet}} - \phi^{\mathrm{parton}}) \in [-\pi, \pi] \label{eq:target_phi}                 \\
  \text{Jet mass fraction:} \quad            & y_3 = \log(m_{\mathrm{jet}} / p_{\mathrm{T}}^{\mathrm{jet}}) \label{eq:target_mass}
\end{align}
At inference the generated residuals are inverted back to ($p_{\mathrm{T}}, \eta, \phi, m$), and a disk clamp keeps the generated jet inside the $\Delta R < 0.2$ matching cone of the source parton. The network does not predict the jet energy. It is reconstructed from the predicted kinematics via $E = \sqrt{p_x^2 + p_y^2 + p_z^2 + m^2}$.

\section{The Bounded CRPS Framework}
\label{sec:CRPS_framework}
A marginal comparison pools all events, so it cannot check a prediction against the event it was conditioned on. We therefore score each test event on its own with the CRPS. For a predictive distribution $F$ and an observed value $y$ it is
\begin{equation}
  \mathrm{CRPS}(F, y) = \int \bigl(F(x) - \mathbf{1}\{y \le x\}\bigr)^2\,\mathrm{d}x = \mathbb{E}_{F}\lvert X - y \rvert - \tfrac{1}{2}\,\mathbb{E}_{F}\lvert X - X' \rvert,
  \label{eq:crps_definition}
\end{equation}
where $X, X' \sim F$ are independent and lower is better. The integral is, up to sign, the uniformly weighted quadratic rule of \citet{Scoring_Rules_for_Continuous_Probability_Distributions}, and the expectation form is given by \citet{gneiting_strictly_2007}. The CRPS is strictly proper for distributions with a finite first moment \citep{gneiting_strictly_2007}, so it penalizes both a wrong location and an over-confident spread. Its absolute value is hard to read without a reference, so we use two. The \emph{floor} is the mean CRPS of the unconditional marginal. This is the best prediction that ignores $c$, and it marks zero skill. The \emph{ceiling} is the best mean CRPS achievable from $c$ (Section~\ref{sec:KNN}). Because lower CRPS is better, the ceiling lies numerically below the floor on a CRPS axis and above it on a skill axis. A model can score worse than the floor, so skill can be negative.

For the marginal of the $N$ test events the expectations in Equation~\ref{eq:crps_definition} can be evaluated exactly. With the test targets sorted as $y_{(1)} \le \dots \le y_{(N)}$, the all-pairs sum reduces to
\begin{equation}
  \mathrm{CRPS}_{\mathrm{floor}} = \frac{1}{N^2} \sum_{i=1}^N (2i - N - 1)y_{(i)}.
  \label{eq:crps_floor_sum}
\end{equation}
This costs $O(N \log N)$ and equals half the Gini mean difference \citep{gini_variabilita_1912} of the empirical distribution of the test targets, with all $N^2$ pairs counted. It is also exactly the uncertainty term $U$ in the CRPS decomposition of \citet{hersbach_decomposition_2000}, whose Eq.~(20) is the same sum written in terms of spacings, so this reference is long established. We use it in place of a Monte Carlo estimate because a plug-in score computed from $m$ marginal draws per event would inflate the floor by a factor $(m+1)/m$ \citep{ferro_effect_2008}. Normalizing a model's mean CRPS, $\mathrm{CRPS}_{\mathrm{cond}}$, by the floor gives the usual skill-score form $1 - A/A_{\text{ref}}$ for an error measure whose perfect score is zero \citep{murphy_skill_1988}, with the sample marginal as reference. Forecast verification calls this a climatological reference \citep{murphy_skill_1988,ferro_effect_2008}, and the resulting quantity, $\mathcal{S} = 1 - \mathrm{CRPS}_{\mathrm{cond}} / \mathrm{CRPS}_{\mathrm{floor}}$, is the Continuous Ranked Probability Skill Score (CRPSS). The Hersbach decomposition writes the CRPS as reliability minus resolution plus uncertainty, which gives $\mathcal{S} \approx (\mathrm{Resol} - \mathrm{Reli})/U$. For a calibrated model $\mathcal{S}$ is therefore the fraction of the floor resolved by $c$, and miscalibration lowers it. The relation is approximate here only because we score samples with the fair estimator of Section~\ref{sec:KNN}.

\textbf{Dependence on the evaluation spectrum.} We use \emph{spectrum} for the distribution $p(c)$ of conditioning states in the test set. Global $\mathcal{S}$ depends on it, because the reference does. \citet{hersbach_decomposition_2000} makes the point for weather, where a sample that pools Norway in January with Spain in March has a very broad climatology and a large uncertainty term. To see the effect on skill, consider a homoscedastic location family $y=\mu(c)+\epsilon$. The conditional CRPS is set by the noise $\epsilon$, while the marginal floor also contains the spread of $\mu(c)$. As $p(c)$ widens the floor grows without limit and global skill tends to $1$, although the noise has not changed. That skill is real, since knowing $c$ does help, but its size says as much about the width of the test set as about the model. We therefore also recompute both the CRPS and the floor within narrow bins of $c$. This is the stratified climatology that \citet{hersbach_decomposition_2000} describes for location and season, with bins of $c$ in their place, and it is the distributional counterpart of the multiple-valued reference of \citet{murphy_skill_1988}. Each bin's reference is then a sampler that knows the bin, so differential skill measures what a model adds beyond the binning variable and no longer depends on how widely that variable is spread. Here we bin in parton $p_{\mathrm{T}}$, and the other components of $c$ stay averaged over. Appendix~\ref{app:sensitivity} shows how the result changes with the number of bins and with the binning variable.

\section{Estimating the Limit with $k$-Nearest-Neighbor Extrapolation}
\label{sec:KNN}
To estimate the ceiling directly from data, we use a differential $k$-nearest-neighbor ($k$-NN) \citep{cover_nearest_1967} extrapolation with four steps. (1) Standardize the continuous conditioning variables and match the discrete flavour exactly. (2) For each query $c_i$, find its $k$ nearest training neighbours for $k \in \{5, 10, 20, 50, 100\}$ and take their true targets as an empirical conditional ensemble. (3) Compute the fair, ensemble-size adjusted CRPS with an $m(m-1)$ kernel \citep{ferro_effect_2008,ferro_fair_2014}, with ensemble size $m = k$, along with the mean neighbourhood radius $r_k$. (4) Fit $(r_k, \text{CRPS}_k)$ across all $k$ by Ordinary Least Squares \citep{james_introduction_2021} and take the $r \to 0$ intercept. The pseudocode is in Appendix~\ref{app:knn_algorithm}.

\textbf{What the estimate guarantees.} The finite-$k$ scores have a clean interpretation. Given the distance to the $(k+1)$-th neighbour, the $k$ retrieved targets are independent draws from the neighbourhood distribution $\bar{F}$, the conditional distribution averaged over that ball. The fair kernel is therefore unbiased for the CRPS of $\bar{F}$ \citep{ferro_fair_2014}. Since the CRPS is proper, $\bar{F}$ cannot beat the true conditional distribution $F$ in expectation, and the excess is $\int (\bar{F} - F)^2\,\mathrm{d}x \ge 0$ \citep{Scoring_Rules_for_Continuous_Probability_Distributions,gneiting_strictly_2007}. So in expectation every finite-$k$ score sits above the Bayes-optimal score for this conditioner. This needs the reference and query events to come from the same distribution, which holds here because both are random splits of one simulated sample.

\begin{figure}[!t]
  \centering
  \includegraphics[width=\linewidth]{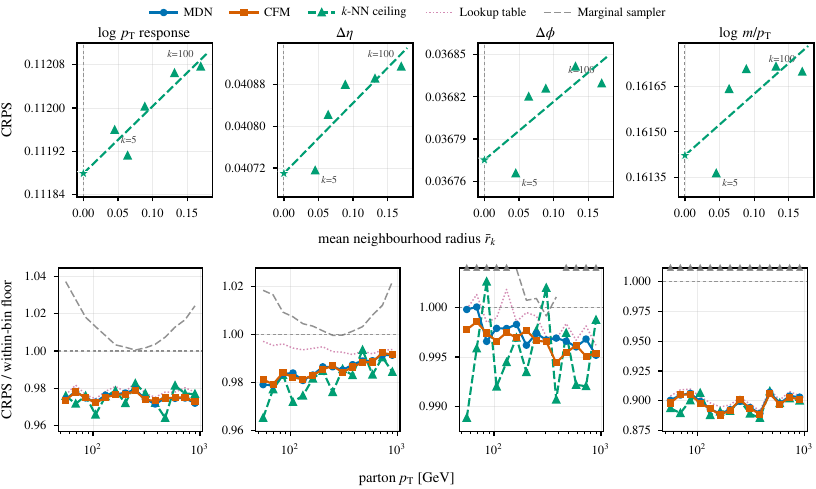}
  \caption{\textbf{Data-driven limits (Seed 1).} Columns are the four targets. \emph{Top.} $k$-NN CRPS against mean neighborhood radius $\bar{r}_k$ for $k \in \{5,10,20,50,100\}$, with the OLS fit and its $r \to 0$ intercept ($\star$). \emph{Bottom.} The differential ceiling over 14 log-spaced parton-$p_{\mathrm{T}}$ bins, with every series (MDN, CFM, $k$-NN ceiling, lookup table and Marginal sampler) scored on the same 200{,}000 test events and read against the floor recomputed inside its own bin. Top-edge triangles mark bins where the Marginal sampler is off scale above. Averaged over the spectrum, both surrogates agree with the $k$-NN estimate to within $\pm 0.005$ (Section~\ref{sec:Target_dependent_saturation}). In single bins the gaps are larger, up to about $0.016$.}
  \label{fig:data_driven_limits}
\end{figure}

No such guarantee holds for the $r \to 0$ intercept, which aims at the Bayes-optimal score itself. If the CRPS really rises with radius and the curve is convex, a straight-line fit lands below the Bayes-optimal score at $r = 0$, so the intercept errs on the low side. That needs both assumptions and is not a theorem. On this benchmark the curve is close to flat, so the intercept simply scatters around the Bayes-optimal score and a model at the limit can land on either side of it. The $k$-NN CRPS barely moves across the ladder (Figure~\ref{fig:data_driven_limits}, top row), the slopes in Table~\ref{tab:knn_slopes_limits} are consistent with zero across splits, and a slope can change sign when the reference subset is redrawn (Appendix~\ref{app:sensitivity}). Neighbourhood blur is small over the whole ladder, so the answer hardly depends on $k$ and the extrapolation changes it very little. We therefore treat the intercept as an estimate of the ceiling and as a consistency check on how much skill the conditioner supports, and we do not claim it as a strict limit.

\section{Empirical Audits}
\label{sec:empirical_audits}
\begin{table}[t]
  \centering
\caption{\textbf{Extrapolated $k$-NN slopes, CRPSS limits, and metric trap inversions.} Every entry is the mean $\pm$ standard deviation over three random data splits (seeds 1--3). The KL columns give the marginal KL divergence of each surrogate divided by that of the condition-blind Marginal sampler on the full test split, so a ratio above $1.0\times$ means the surrogate is worse. $\mathrm{d}\mathrm{CRPS}/\mathrm{d}r$ is the OLS slope of the $k$-NN CRPS on neighbourhood radius, in target units per unit of standardized radius, and values consistent with zero mean the estimate barely depends on $k$. $S_{\mathrm{global}}$ and $S_{\mathrm{diff}}$ are the skill of the $r \to 0$ ceiling against the global floor and against the floor recomputed within each of 14 parton-$p_{\mathrm{T}}$ bins (median over bins). LUT Rec.\ is $S_{\mathrm{LUT}}/S_{\mathrm{CFM}}$, the lookup-table skill as a fraction of the CFM skill. CRPS quantities use the first 200{,}000 test events of each split with $m = 200$ samples.}
\label{tab:knn_slopes_limits}
\vspace{0.4em}
\small\setlength{\tabcolsep}{3pt}
\begin{tabular}{lcccccc}
  \toprule
  \textbf{Target}                 & \textbf{KL MDN /}    & \textbf{KL CFM /}     & \textbf{$\mathrm{d}\mathrm{CRPS}/\mathrm{d}r$} & \textbf{$S_{\mathrm{global}}$} & \textbf{$S_{\mathrm{diff}}$} & \textbf{LUT}     \\
                                  & \textbf{Marginal}    & \textbf{Marginal}     & ($\times 10^{-3}$)                             &                                &                              & \textbf{Rec.}    \\
  \midrule
  $\log p_{\mathrm{T}}$ response  & $5.5 \pm 1.7\times$  & $0.90 \pm 0.21\times$ & $+0.51 \pm 0.64$                               & $3.5 \pm 0.1\%$                & $2.3 \pm 0.1\%$              & $90.8 \pm 0.5\%$ \\
  $\log m/p_{\mathrm{T}}$         & $4.3 \pm 0.6\times$  & $1.11 \pm 0.19\times$ & $+0.12 \pm 2.95$                               & $28.7 \pm 0.1\%$               & $10.1 \pm 0.4\%$             & $98.8 \pm 0.1\%$ \\
  $\Delta\eta$                    & $16.7 \pm 1.9\times$ & $4.1 \pm 0.3\times$   & $+0.57 \pm 0.69$                               & $2.2 \pm 0.3\%$                & $1.5 \pm 0.2\%$              & $60.8 \pm 1.4\%$ \\
  $\Delta\phi$                    & $19.0 \pm 2.9\times$ & $6.4 \pm 0.7\times$   & $+0.19 \pm 0.23$                               & $1.3 \pm 0.1\%$                & $0.4 \pm 0.2\%$              & $84.6 \pm 1.1\%$ \\
  \bottomrule
\end{tabular}
\end{table}

\subsection{The Metric Trap}
\label{sec:metric_trap}
The condition-blind Marginal sampler defeats the correctly conditioned MDN on 8 of 8 target-divergence combinations at each of three random data splits, by margins of $3.65\text{--}22.39\times$ under KL and $3.75\text{--}18.02\times$ under $W_1$. Table~\ref{tab:knn_slopes_limits} gives the KL half per target as a mean over splits, while the ranges here are per-split extremes. The Marginal sampler's KL is at the histogram sampling-noise level (about $2\times10^{-4}$, against $1\text{--}4\times10^{-3}$ for the MDN), so the large ratios partly reflect how small the sampler's KL is. The CFM is closer. On momentum response and jet mass it changes sides with the sampler between splits, while on $\Delta\eta$ and $\Delta\phi$ the sampler wins at every split. The MDN reproduces the marginals worse than the CFM on every target and split, under both divergences, yet the two are almost indistinguishable on the per-event CRPS, where their spectrum-averaged ratios differ by at most $0.0007$ (Section~\ref{sec:Target_dependent_saturation}). On this benchmark a mixture head therefore matches a flow event by event but not on the marginals, which bears on the open question raised by \citet{baattrup_pointwise_2026} of whether mixture heads can match flows in spectrum fidelity on realistic problems.

\subsection{Target-Dependent Saturation}
\label{sec:Target_dependent_saturation}
Measured against the $k$-NN ceiling (Figure~\ref{fig:data_driven_limits}), both surrogates agree with the estimated limit. On every target and data split, the spectrum-averaged CRPS ratios of both learned surrogates sit within $\pm 0.005$ of the global $k$-NN intercept ratio, scored on the same events, and they fall on both sides of it. This holds when the reference subset is redrawn and when the full training set is used as reference (Appendix~\ref{app:sensitivity}).

To test whether that limit reflects model capacity or the conditioning state itself, we replace the networks with a parameter-free lookup table. The conditioner is reduced to four binned axes, $40$ bins in $\log p_{\mathrm{T}}$, $12$ in $|\eta|$, $6$ in \texttt{nTrueInt}, and the $6$ exact-match flavour strata, giving $40 \times 12 \times 6 \times 6 = 17{,}280$ cells, and each query is answered by resampling the training targets stored in its cell. Bin edges are training-set quantiles, so each continuous axis is equally populated. We do not bin $\log(\cosh\eta)$ because it is a deterministic function of $|\eta|$. Cells holding fewer than $20$ training rows back off to the flavour marginal, and then to the global marginal. Averaged over the three data splits, the table recovers $98.8\%$ of the CFM's spectrum-averaged conditional skill on jet mass and $84.6\%$ on $\Delta\phi$, though only $60.8\%$ on $\Delta\eta$, with no gradients and no learned parameters. Capacity is therefore unlikely to be the binding constraint, except possibly on $\Delta\eta$.

The gap between $S_{\mathrm{global}}$ and $S_{\mathrm{diff}}$ in Table~\ref{tab:knn_slopes_limits} shows why we also recompute the floor locally. If the $r \to 0$ $k$-NN ceiling is scored first against the global floor and then against the floor recomputed inside each $p_{\mathrm{T}}$ bin, its skill on jet mass falls from $28.7\%$ to $10.1\%$, a relative drop of $64.7\%$ (mean over splits, per-split range $63.6\text{--}66.1\%$). The difference is real information, since jet mass fraction changes strongly with $p_{\mathrm{T}}$ across the evaluated range, but its size depends on the width of the spectrum.

\subsection{Likelihood and CRPS Near the Limit}
\label{sec:Metric_decoupling}
\begin{figure}[!t]
  \centering
  \includegraphics{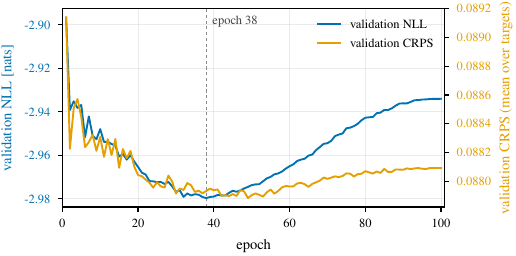}
  \caption{\textbf{Likelihood and CRPS past the best epoch (Seed 1, MDN).} Validation NLL (left axis) and validation CRPS averaged over the four targets (right axis). The rule marks epoch 38, the checkpointed best-validation-loss epoch. Past it, the NLL worsens by 1.53\% and the CRPS by only 0.18\%. The two axes are scaled independently, so the CRPS rise looks larger than it is. The CFM is omitted because it does not overfit.}
  \label{fig:overfit_decoupling}
\end{figure}

The two scores respond differently to overfitting (Figure~\ref{fig:overfit_decoupling}). We anchor at the best validation Negative Log-Likelihood (NLL), which for seed 1 falls at epoch 38. Training the MDN for 62 further epochs worsens validation NLL by 0.046 nats (1.53\%). Validation CRPS, tracked on a fixed 20{,}000-event subset of the same split with $m = 20$ samples, worsens by only 0.18\% over the same epochs, and seeds 2 and 3 show the same pattern. The NLL is itself a strictly proper scoring rule \citep{gneiting_strictly_2007}, so neither score has failed here and the two simply weight errors differently. The percentages cannot be compared directly, because the NLL has no natural zero and because it scores the joint 4D density while the CRPS here is a mean of 1D scores. In practice a flat CRPS does not rule out overfitting, and we would monitor it alongside the likelihood or a calibration test. \citet{baattrup_pointwise_2026} find the same blind spot on top-quark reconstruction, where coverage separates two flows with comparable CRPS.

\section{Limitations}
\label{sec:limitations}
The ceiling is an estimate and does not prove saturation. It is computed from the continuous conditioner and flavour, and Appendix~\ref{app:sensitivity} shows that the $\pm 0.005$ agreement holds when the reference subset is redrawn and when the full training set is used. The surrogates also receive $\phi$, which the neighbourhood search does not, and this explains the one consistent gap we see. On momentum response they sit about 0.0014 below the $k$-NN score averaged over $k \ge 20$ at every split, and adding $\phi$ to the search closes that gap. Their offset from the extrapolated intercept is noisier and changes sign between splits. On the angular targets $\pm 0.005$ is a sizeable fraction of the available skill, so saturation there is only weakly constrained. The OLS standard errors in Algorithm~\ref{alg:knn_extrapolation} treat the five ladder points as independent, which they are not, since they share query events and nested neighbourhoods. They should be read as indicative only, and the redraw scatter in Appendix~\ref{app:sensitivity} is a better guide.

All scores are per-target $1\text{D}$ scores and do not test correlations between targets. Multivariate proper scores such as the energy score \citep{gneiting_strictly_2007} and joint tests such as conditional classifiers are still needed, as are marginal checks, because a per-event score does not measure the population-level spectrum \citep{baattrup_pointwise_2026}.

Differential skill depends on what is binned. The number of parton $p_{\mathrm{T}}$ bins hardly matters, but binning in $|\eta|$ or jointly in $p_{\mathrm{T}}$ and flavour gives very different values (Appendix~\ref{app:sensitivity}), so a differential skill means little unless its binning is stated. Momentum response is also weakly conditional by design. We use calibrated jets, whose energy corrections are built to flatten the response, so low skill on that target is expected. Our numbers are specific to this task and should not be carried over to other surrogates without repeating the procedure.

\section{Conclusion}
Marginal divergences alone can reward conditional collapse, as a condition-blind sampler shows on this benchmark. A CRPS skill score with the exact marginal reference is easier to interpret, but global skill depends on the breadth of the evaluation spectrum, so it should be reported alongside a differential skill that names its binning. A $k$-NN extrapolation then estimates how much skill the conditioner supports. Two widely used conditional density estimators agree with that estimate in parton-to-jet simulation, and a binned lookup table with no learned parameters recovers most of the same skill on most targets. This suggests the limit here is set mainly by the conditioning state and not by model capacity.

\section*{Acknowledgements}
This work was carried out by C.D. as an undergraduate research project under the supervision of S.D. It uses simulated data released through the CERN Open Data Portal \citep{cms_collaboration_qcd_pt-15to7000_tunecp5_flat2018_13tev_pythia8runiisummer20ul16nanoaodv9-106x_mcrun2_asymptotic_v17-v1nanoaodsim_2024}. The authors would
like to acknowledge the efforts put in by the CMS Collaboration to release the simulation samples used in this work.
%The results are the authors' own and are not endorsed by the CMS Collaboration or CERN.

\section*{Description of AI Use}
Generative AI was used as a collaborative programming assistant to help write code snippets, and debug scripts, as well as for LaTeX formatting and copy-editing. To verify the accuracy of any generated code or text, we established a fully reproducible, end-to-end evaluation pipeline with provenance checks and tests. All empirical metrics and physics boundaries were directly computed from raw data and manually validated to ensure they adhere to expected mathematical constraints (such as the exact CRPS floors and $k$-NN ceilings). The core scientific methodology, claims, physical interpretations, mathematical derivations, and final conclusions were the authors' own and were rigorously verified.
\clearpage

\bibliographystyle{unsrtnat}
\bibliography{references}

\clearpage
%%%%%%%%%%%%%%%%%%%%%%%%%%%%%%%%%%%%%%%%%%%%%%%%%%%%%%%%%%%%
%%%%%%%%%%%%%%%%%%%%%%%%%%%%%%%%%%%%%%%%%%%%%%%%%%%%%%%%%%%%
\appendix
\counterwithin{figure}{section}
\counterwithin{table}{section}

\section{Model Architectures and Training Setup}
\label{app:model_and_training}

For a like-for-like comparison, both the Mixture Density Network (MDN) \citep{bishop_mixture_1994} and Conditional Flow Matching (CFM) \citep{lipman_flow_2023} surrogates are built on the same backbone, a 6-layer Multilayer Perceptron (MLP) with a hidden width of 512 dimensions and SiLU activations \citep{ELFWING20183}. The models are not matched in total size. With its output heads the MDN has 1.95 million parameters, and the CFM has 1.43 million including its time embedding. Both models are trained for 100 epochs using a batch size of 4,096 on the same data splits, and the seed of a split also sets the initialization and the batch order. Optimization uses AdamW \citep{Loshchilov2017DecoupledWD} (learning rate $3 \times 10^{-4}$, weight decay $1 \times 10^{-5}$) with a linear warmup over the first 2{,}000 steps and cosine annealing \citep{Loshchilov2016SGDRSG}, and gradients are clipped at norm 1.0. Each model is evaluated at the epoch with the best validation loss, which is the NLL for the MDN and the velocity MSE for the CFM.

\subsection{Mixture Density Network (MDN)}
The MDN outputs parameters for an 8-component multivariate Gaussian mixture. The shared MLP backbone branches into three separate 3-layer MLP heads (hidden width 256) that predict the mixture weights ($\pi$), means ($\mu$), and full lower-triangular Cholesky covariance factors ($L$) for each component. To prevent variance collapse during training, the Cholesky diagonal uses a softplus activation paired with a hard epsilon floor.

The MDN is optimized on the exact conditional Negative Log-Likelihood (NLL) with a variance calibration penalty. This penalty (weighted by $\lambda=0.1$) acts on the $\pi$-weighted mixture moments, pulling the batch mean of the normalized squared residual toward the target dimensionality ($D=4$). To stabilize early training, weight decay is disabled for the output heads, and a cosine temperature annealing schedule is applied to the $\pi$ logits over the first half of training. During inference, the model selects the active mixture component for each event using the Gumbel-max trick \citep{maddison2014sampling}, followed by standard reparameterization sampling using the selected component's mean and Cholesky factor.

\subsection{Conditional Flow Matching (CFM)}
The CFM model acts as a continuous normalizing flow \citep{chen2018neural}, parameterized by a velocity field network $v_\theta(x_t, t, c)$. The integration time $t \sim \text{Uniform}(0, 1)$ is projected via a 128-dimensional sinusoidal time embedding network. This time embedding is concatenated directly with the noisy state $x_t$ and the 26-dimensional conditioner $c$ before passing through the MLP backbone.

The CFM is trained using an independent noise coupling strategy. The target state $x_1$ is drawn from the dataset, and the base state $x_0 \sim \mathcal{N}(\mu, \sigma^2)$ is drawn from a Gaussian noise prior explicitly moment-matched to the mean and variance of the empirical training targets. The model is optimized using Mean Squared Error (MSE) between the network's predicted velocity and the target velocity vector $v_{\text{target}} = x_1 - x_0$. At evaluation, samples are drawn by integrating the learned velocity field with a midpoint solver over 100 steps, for every CFM score reported here. Only the per-epoch validation check during training uses 20 steps. Table~\ref{tab:knn_slopes_limits}, Figure~\ref{fig:data_driven_limits} and Figure~\ref{fig:knn_3seeds} score both surrogates on the first 200{,}000 test events of each split with $m = 200$ samples per event. The conditioner ladder (Figure~\ref{fig:conditioner_ladder}) and the generator-level oracle (Figure~\ref{fig:genjet_oracle}) compute the CRPS on the first 50{,}000 test events with $m = 50$.

\section{Robustness to the Matching Cone}
\label{app:cone_narrowing}

The angular channels ($\Delta\eta$ and $\Delta\phi$) have very little conditional skill (Section~\ref{sec:Target_dependent_saturation}). Target jets are matched to partons with a $\Delta R < 0.2$ cone, which truncates the tails of the angular targets. One might therefore ask whether the low skill is a real property of the problem or an artifact of this selection boundary.

To test this, we evaluated the differential $k$-NN ceiling on narrower cone subsets ($\Delta R < 0.15$ and $\Delta R < 0.10$) at each of the three data splits.
Halving the matching cone to $\Delta R < 0.10$ narrowed the raw target standard deviations, averaged over the three splits, by $1.837 \pm 0.006\times$ (from 0.0746 to 0.0406) for $\Delta\eta$, and by $1.799 \pm 0.003\times$ (from 0.0677 to 0.0376) for $\Delta\phi$. If the low conditional skill were a boundary artifact, tightening the boundary this much should change the normalized CRPS ratio.

\begin{figure}[!t]
  \centering
  \includegraphics{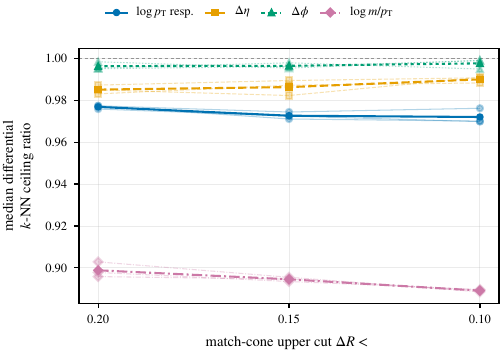}
  \caption{\textbf{Cone-narrowing robustness check.} The median differential $k$-NN ceiling ratio evaluated across tightening $\Delta R$ cuts. Faint lines show the three data splits and bold lines their mean. While tightening the cone to $0.10$ drastically reduces the raw standard deviation of the angular targets, the normalized conditional skill remains nearly flat, so the low predictability of $\Delta\eta$ and $\Delta\phi$ is not a boundary artifact.}
  \label{fig:cone_narrowing}
\end{figure}

Instead, as shown in Figure~\ref{fig:cone_narrowing}, the median $k$-NN ceiling ratios changed little, moving on average by $+0.0049 \pm 0.0013$ for $\Delta\eta$ (0.9852 $\to$ 0.9901) and $+0.0013 \pm 0.0023$ for $\Delta\phi$ (0.9964 $\to$ 0.9977). For $\Delta\eta$ this shift was toward \emph{less} apparent skill at every split, and for $\Delta\phi$ the mean shift points the same way, though it is consistent with zero across splits. The ratios change little under a large reduction in raw variance, so the low angular skill is not an artifact of the cone boundary.

\section{Stage-Specific Ceilings \& The Generator-Level Oracle}
\label{app:stage_specific_ceiling}
Applying the differential $k$-NN ceiling across intermediate simulation stages compares the stochasticity of the stage before generator-level jet formation with that of detector response, without training a network.

The standard surrogate task (parton2reco) maps directly from the hard-scatter parton to the reconstructed detector-level jet. We compute the model-free $k$-NN ceiling independently for the parton2genjet stage (parton shower, hadronization and underlying event) and the genjet2reco stage (detector simulation) across common parton $p_T$ bins. All three stages use the same common event set, for both the reference and the query events.

\begin{figure}[!t]
  \centering
  \includegraphics[width=\linewidth]{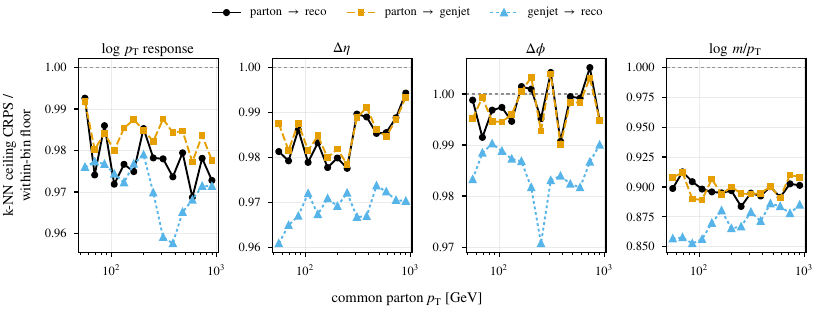}
  \caption{\textbf{Model-free differential stage ceilings.} Evaluated via $k$-NN extrapolation across common $p_{\mathrm{T}}$ bins (Seed 1). By computing the ceiling for the direct mapping ($\text{parton} \to \text{reco}$) versus its intermediate physical stages ($\text{parton} \to \text{genjet}$, $\text{genjet} \to \text{reco}$), we compare the stochasticity added before and after generator-level jet formation.}
  \label{fig:app_stage_ceilings}
\end{figure}

As shown in Figure~\ref{fig:app_stage_ceilings}, the detector response stage (genjet2reco) has a lower differential ceiling ratio across all targets than the parton2genjet stage, which means more skill against its own floor. For example, on seed 1 (the split shown in Figure~\ref{fig:app_stage_ceilings}) the median differential ceiling ratio for jet mass fraction sits at 0.871 for the detector stage, compared to 0.897 for parton2genjet. This comparison is descriptive only, for the reason given next.

Changing the conditioning stage, for example from a parton input to a genjet input, also redefines the target, and the exact unconditional floor shifts by up to $6.4\times$ on angular targets. Floor-normalized ratios therefore cannot be compared across stages unless the target distribution is the same.

\begin{figure}[!t]
  \centering
  \includegraphics{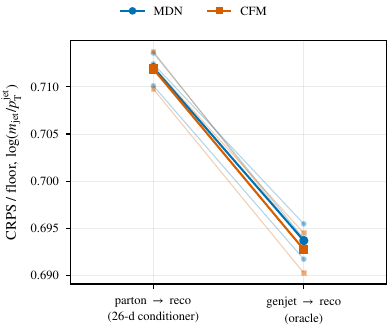}
  \caption{\textbf{Generator-level oracle, jet-mass channel only (seeds 1--3).} Floor-normalized CRPS for the standard \texttt{parton2reco} surrogate and for the same architecture handed the unobservable generator-level jet, both scored on the first 50{,}000 events of the identical common event set with $m = 50$. Faint lines show the three data splits and bold lines their mean. \texttt{log\_mass\_frac} is the only target shown because it is the only one whose exact unconditional floor is identical (to five decimal places, 0.22580) for the two conditioning stages. On the angular targets the floor shifts by up to $6.4\times$ and the two ratios share no axis. Lower is better, so the drop is the skill gained from the generator-level jet's 3-vector and flavour.}
  \label{fig:genjet_oracle}
\end{figure}

To compare stages without these floor shifts, we used a generator-level oracle. On the jet mass target, the exact unconditional floors for both tasks match to five decimal places at every split (0.22580 on average). There, replacing the parton conditioner with the kinematics and flavour of the unobservable generator-level jet gave a relative gain in skill of $6.4 \pm 0.5\%$ (MDN) and $6.6 \pm 0.5\%$ (CFM) on a common event set, averaged over the three splits. The generator-level jet mass is not part of this conditioner, so the gain measures what its 3-vector and flavour add, and it does not measure the total information lost before the detector. Figure~\ref{fig:genjet_oracle} shows this difference directly.

\section{Full 3-Seed $k$-NN Extrapolations}
\label{app:three_seeds}
As outlined in Section~\ref{sec:KNN}, the $r \to 0$ intercept is an estimate of the best achievable score rather than a strict bound, and we use it as a consistency check.

\begin{figure}[!t]
  \centering
  \includegraphics[width=\linewidth]{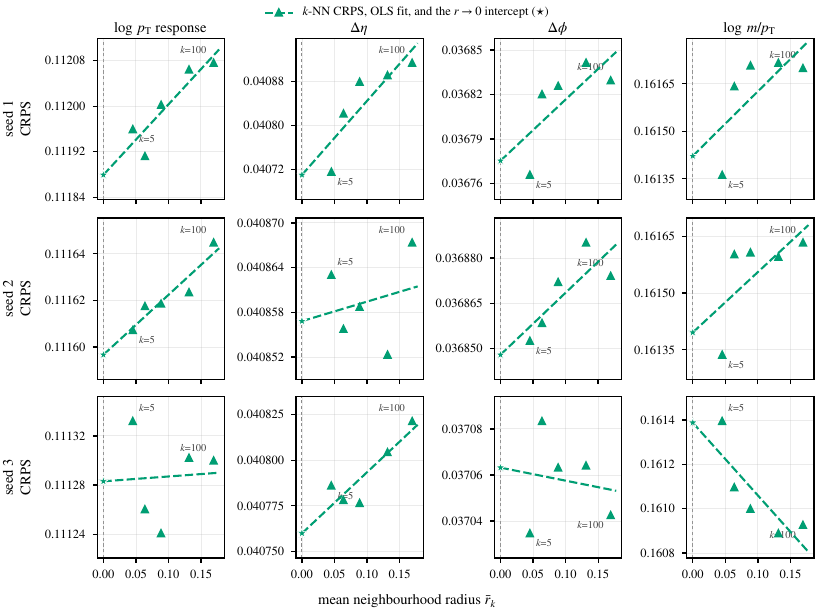}
  \caption{\textbf{Full 3-seed $k$-NN extrapolations.} $k$-NN extrapolation to zero mean neighborhood radius ($r \to 0$) evaluated across three random data splits (Seeds 1, 2, and 3, top to bottom). Columns are the four targets. The intercept ($\star$) estimates the Bayes-optimal CRPS. The slope is positive at all three splits for momentum response and $\Delta\eta$ (the two leftmost columns). Each panel is scaled to its own fit. The exact unconditional marginal floor lies above every panel here and enters Table~\ref{tab:knn_slopes_limits} as the denominator of $S_{\mathrm{global}}$.}
  \label{fig:knn_3seeds}
\end{figure}

To show the radius dependence behind the slopes in Table~\ref{tab:knn_slopes_limits}, Figure~\ref{fig:knn_3seeds} gives the full $k$-NN extrapolation across all three data splits (Seeds 1, 2, and 3). The slope is positive at all three splits for momentum response and $\Delta\eta$, and changes sign between splits for jet mass and $\Delta\phi$. These signs should not be over-read, because the curves are nearly flat relative to the floor in every case and many of the signs change when the reference subset is redrawn (Appendix~\ref{app:sensitivity}). The surrogates lie on both sides of the intercept, within $\pm 0.005$ in ratio.

\section{Conditioner Ablation Ladder}
\label{app:conditioner_ladder}
\begin{figure}[!t]
  \centering
  \includegraphics[width=\linewidth]{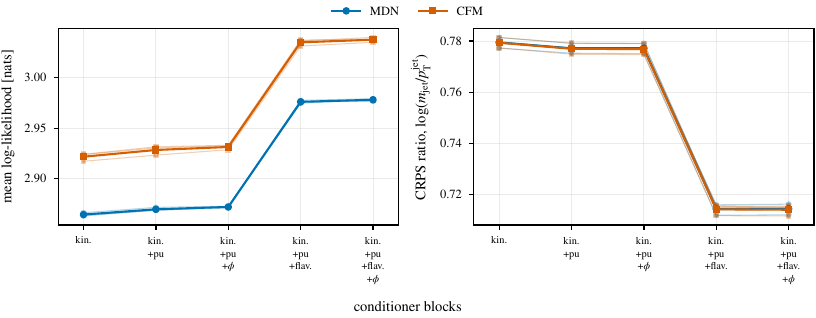}
  \caption{\textbf{Conditioner ablation ladder (seeds 1--3).} Faint lines show individual data splits and bold lines their mean. Mean conditional log-likelihood (left) and the floor-normalized CRPS on \texttt{log\_mass\_frac} (right) against the conditioner blocks supplied to the network. \texttt{kin.}\ is the three standardized parton kinematics, \texttt{pu} the pileup scalar, \texttt{flav.}\ the six-way one-hot flavor identity, and $\phi$ the 16 Fourier azimuth features. Higher is better on the left, lower on the right. The MDN log-likelihood uses the full test split and the CFM log-likelihood the first 100{,}000 test events.}
  \label{fig:conditioner_ladder}
\end{figure}

The 26-dimensional conditioning state of Section~\ref{app:physics_task} is built from four blocks, and they do not contribute equally. Figure~\ref{fig:conditioner_ladder} retrains both surrogates on each rung of the ladder, ordered by information content, and reports the mean conditional log-likelihood alongside the CRPS ratio on the jet-mass channel, the target with the most conditional information. Each rung is trained at three random data splits (seeds 1--3). Faint lines show the individual splits and bold lines their mean. The CRPS ratio is scored on the first 50{,}000 test events of each split with $m = 50$ samples. The log-likelihood is exact for the MDN and uses the full test split. For the CFM it is obtained by integrating the exact divergence along the flow, extrapolated from 100 and 200 midpoint steps, on the first 100{,}000 test events.

Discrete flavor identity dominates. Adding it is worth roughly an order of magnitude more likelihood than pileup and the Fourier azimuth combined, and it accounts for essentially the whole of the jet-mass CRPS improvement across the ladder. The two model families agree on the ordering of the rungs and on the size of the flavour step, which is 0.106 nats for both. The remaining steps are small for both models, so the ladder describes the conditioning state and not either density estimator.

\section{Sensitivity Checks}
\label{app:sensitivity}
These checks reuse the stored samples of the trained models and run on the same 200{,}000 test events as Table~\ref{tab:knn_slopes_limits}, at all three data splits.

\textbf{Binning.} Table~\ref{tab:binning} repeats the differential evaluation under five binnings. The number of parton $p_{\mathrm{T}}$ bins has little effect. Going from 14 to 28 bins moves every entry by less than $0.2$ percentage points, and at 7 bins only jet mass moves noticeably, by about half a percentage point, which is expected when the bins are this wide.

The binning variable matters much more. Binning in $|\eta|$ leaves almost all of the global skill on jet mass in place ($27.9\%$ against $S_{\mathrm{global}} = 28.7\%$), which shows that $|\eta|$ carries little information about jet mass fraction. Binning jointly in $p_{\mathrm{T}}$ and flavour removes most of it. The ceiling keeps about $3\%$ of skill on jet mass and about $0.1\%$ on momentum response and $\Delta\phi$, while $\Delta\eta$ keeps about $1\%$. Most of the conditional information in these targets is therefore carried by parton $p_{\mathrm{T}}$ and flavour, in line with the ablation ladder of Appendix~\ref{app:conditioner_ladder}. In every row of the table the surrogates stay within about $0.2$ percentage points of the ceiling, so their agreement with it does not depend on this choice.

\begin{table}[h]
  \centering
  \caption{\textbf{Differential skill under different binnings.} Skill in percent, taken as the median over bins, with mean $\pm$ standard deviation over three data splits. The $k$-NN rows use the $r \to 0$ ceiling. Every bin holds at least 200 test events, and the median $p_{\mathrm{T}} \times$ flavour bin holds about 750. Because those bins are small, the floor here is corrected from the $N^2$ divisor to $N(N-1)$ in each bin. This moves the 14-bin values by about $0.01$ percentage points, so they round to the $S_{\mathrm{diff}}$ column of Table~\ref{tab:knn_slopes_limits}.}
  \label{tab:binning}
  \vspace{0.4em}
  \small\setlength{\tabcolsep}{4pt}
  \begin{tabular}{llcccc}
    \toprule
    & \textbf{Binning} & $\log p_{\mathrm{T}}$ response & $\log m/p_{\mathrm{T}}$ & $\Delta\eta$ & $\Delta\phi$ \\
    \midrule
    $k$-NN & $p_{\mathrm{T}}$, 7 bins & $2.39 \pm 0.15$ & $10.68 \pm 0.23$ & $1.58 \pm 0.37$ & $0.42 \pm 0.09$ \\
           & $p_{\mathrm{T}}$, 14 bins & $2.32 \pm 0.09$ & $10.13 \pm 0.36$ & $1.49 \pm 0.21$ & $0.37 \pm 0.18$ \\
           & $p_{\mathrm{T}}$, 28 bins & $2.35 \pm 0.10$ & $10.16 \pm 0.44$ & $1.43 \pm 0.28$ & $0.33 \pm 0.12$ \\
           & $|\eta|$, 14 bins & $3.43 \pm 0.05$ & $27.85 \pm 0.14$ & $1.32 \pm 0.34$ & $1.24 \pm 0.15$ \\
           & $p_{\mathrm{T}} \times$ flavour, 84 bins & $0.13 \pm 0.08$ & $2.95 \pm 0.18$ & $1.09 \pm 0.12$ & $0.13 \pm 0.18$ \\
    \midrule
    MDN & $p_{\mathrm{T}}$, 7 bins & $2.54 \pm 0.06$ & $10.68 \pm 0.25$ & $1.43 \pm 0.05$ & $0.29 \pm 0.03$ \\
        & $p_{\mathrm{T}}$, 14 bins & $2.49 \pm 0.05$ & $10.17 \pm 0.09$ & $1.45 \pm 0.03$ & $0.28 \pm 0.03$ \\
        & $p_{\mathrm{T}}$, 28 bins & $2.48 \pm 0.04$ & $10.05 \pm 0.07$ & $1.44 \pm 0.02$ & $0.29 \pm 0.02$ \\
        & $|\eta|$, 14 bins & $3.32 \pm 0.06$ & $27.89 \pm 0.20$ & $1.20 \pm 0.11$ & $1.15 \pm 0.05$ \\
        & $p_{\mathrm{T}} \times$ flavour, 84 bins & $0.30 \pm 0.03$ & $3.07 \pm 0.14$ & $0.91 \pm 0.12$ & $-0.02 \pm 0.02$ \\
    \midrule
    CFM & $p_{\mathrm{T}}$, 7 bins & $2.54 \pm 0.09$ & $10.75 \pm 0.19$ & $1.51 \pm 0.04$ & $0.32 \pm 0.02$ \\
        & $p_{\mathrm{T}}$, 14 bins & $2.51 \pm 0.03$ & $10.29 \pm 0.07$ & $1.52 \pm 0.00$ & $0.33 \pm 0.01$ \\
        & $p_{\mathrm{T}}$, 28 bins & $2.49 \pm 0.06$ & $10.11 \pm 0.03$ & $1.46 \pm 0.02$ & $0.33 \pm 0.02$ \\
        & $|\eta|$, 14 bins & $3.33 \pm 0.05$ & $27.94 \pm 0.19$ & $1.24 \pm 0.09$ & $1.25 \pm 0.04$ \\
        & $p_{\mathrm{T}} \times$ flavour, 84 bins & $0.31 \pm 0.02$ & $3.16 \pm 0.08$ & $0.86 \pm 0.13$ & $0.03 \pm 0.03$ \\
    \bottomrule
  \end{tabular}
\end{table}

\textbf{Reference subset.} The main results use a random 400{,}000 reference events per flavour. We redrew that subset 12 times at each split. The intercept ratio scatters with a standard deviation of $0.0008\text{--}0.0013$, which has a median of $2.5$ times the OLS standard error and reaches $24$ times it. All 312 comparisons between a model and an intercept (13 draws, three splits, four targets, two models) stay within $\pm 0.005$, the largest difference is $0.0041$, and the models fall below the intercept in $54\%$ of them. The slope is not stable under redrawing. It is positive in $42\text{--}67\%$ of redraws on momentum response, $25\text{--}58\%$ on jet mass and $25\text{--}75\%$ on $\Delta\phi$, depending on the split. Only $\Delta\eta$ leans clearly positive, at $58\text{--}100\%$.

\textbf{Full reference.} We also removed the cap and used all of the roughly 8.6 million training events as reference, which shrinks the neighbourhood radii by about a factor of two. The intercept ratio moves by at most $0.0027$ from the mean of the redraws. All 24 comparisons stay within $\pm 0.005$, with a largest difference of $0.0022$, so the cap does not limit the estimate.

\textbf{Azimuth.} The surrogates receive $\phi$ and the neighbourhood search does not. To see whether this matters we repeated the full-reference search with the first Fourier pair $(\sin\phi, \cos\phi)$ added to the search space, scaled to unit variance. We compare the ratio averaged over $k \ge 20$, because the noise on the intercept is as large as the effect we are looking for. On momentum response, including $\phi$ lowers the $k$-NN score by $0.0017 \pm 0.0006$. The surrogates sit $0.0014$ below the search without $\phi$ and $0.0003$ above the search with it, so $\phi$ accounts for the gap. Using $k = 100$ alone gives the same picture ($0.0015$ lower, and the surrogates then $0.0002$ above). On average, jet mass and $\Delta\phi$ move by less than $0.0005$. On $\Delta\eta$ the $k$-NN score rises by $0.0014 \pm 0.0008$, which is consistent with the larger neighbourhood, since the radii in this search are three to five times wider at $k \ge 20$.

\section{$k$-NN Extrapolation Algorithm}
\label{app:knn_algorithm}

Algorithm~\ref{alg:knn_extrapolation} gives the procedure used to compute the data-driven ceiling.
The idea is simple. For a query event, the true targets of its $k$ most similar training events give an empirical picture of the conditional distribution. That picture is blurred, because the $k$ neighbours are not at identical conditions, so the score it earns is slightly too high. A smaller neighbourhood is sharper but has fewer events. Measuring the score at several $k$ and extrapolating the trend to zero radius estimates the score at zero blur, which is the best any model with the same conditioning information could achieve.

Three details in the implementation matter enough to state explicitly.

\begin{itemize}
  \item \textbf{Neighbours are used as they are.} The $k$ retrieved targets are scored directly as a $k$-member ensemble. The expected value of the $m(m-1)$ kernel \citep{ferro_fair_2014} does not depend on ensemble size \citep{ferro_effect_2008}, so drawing extra copies with replacement adds nothing and does harm. A resampled ensemble makes the kernel score the $k$-atom empirical distribution instead of the neighbourhood distribution. This adds a positive bias of about $\mathbb{E}\lvert X - X' \rvert / (2k)$, which is largest at small $k$, so it pushes the score up there and can flip the sign of the trend.
  \item \textbf{A query is excluded from its own neighbourhood by event identity, not by distance.} If a query is also in the reference set, it retrieves itself and one ensemble member is a perfect prediction. This lowers the score by $40\%$ at $k=5$ (measured by querying reference events against their own search structure), in exactly the direction that would fake a model beating the ceiling. Distance cannot be used for this test, because NanoAOD stores $p_{\mathrm{T}}$ and $\eta$ at reduced precision. Different jets can therefore sit at zero distance from one another ($2\%$ of queries against the capped reference set), and discarding them would throw away the best neighbours available.
  \item \textbf{Flavour is matched exactly.} A separate search structure is built per flavour, so no neighbour is ever borrowed across flavours. Distances are computed over the four standardized continuous features (the three kinematic ones and pileup) and not the full $26$-D conditioner, so the azimuth features are not used in the neighbourhood search.
\end{itemize}

\begin{algorithm}[h]
  \caption{Data-Driven Performance Ceiling via $k$-NN Extrapolation}
  \label{alg:knn_extrapolation}
  \begin{algorithmic}[1]
    \Require Disjoint reference and query sets of events, each carrying a standardized continuous conditioner $z$, a flavour $f$, targets $y \in \mathbb{R}^{D}$, and a unique event id
    \Require Neighbourhood ladder $\mathcal{K} = \{5, 10, 20, 50, 100\}$
    \Ensure Per-target ceiling $\mathrm{CRPS}^{\star}$, slope $\mathrm{d}\mathrm{CRPS}/\mathrm{d}r$, and their standard errors
    \Statex
    \Statex \textbf{Stage 1: Stratify and standardize}
    \State Standardize the continuous conditioner using training-set statistics.
    \For{each flavour $f$}
    \State Collect the reference events of that flavour (capped at $M_f$ events) and build a KD-tree on their $z$.
    \EndFor
    \Statex
    \Statex \textbf{Stage 2: Sweep the neighbourhood ladder}
    \For{$k \in \mathcal{K}$} \Comment{$k \ge 2$ is required by the fair estimator}
    \For{each query event $i$}
    \State Retrieve its $k$ nearest reference events \emph{within its own flavour}, excluding any neighbour whose event id equals $i$'s.
    \State Take their true targets as the conditional ensemble $S_i$ (used as-is, never resampled).
    \State $\mathrm{crps}_{i} \gets \Call{FairCRPS}{y_i, S_i}$, computed per target dimension.
    \EndFor
    \State $\mathrm{CRPS}_k \gets$ mean of $\mathrm{crps}_i$ over queries.
    \State $\bar{r}_k \gets$ mean distance to the $k$-th (farthest retained) neighbour.
    \EndFor
    \Statex
    \Statex \textbf{Stage 3: Extrapolate to zero radius}
    \For{each target dimension}
    \State Fit $\mathrm{CRPS}_k = \beta\,\bar{r}_k + \alpha$ by ordinary least squares over the $n = |\mathcal{K}|$ points.
    \State $\mathrm{CRPS}^{\star} \gets \alpha$ (the $r \to 0$ intercept); \quad $\mathrm{d}\mathrm{CRPS}/\mathrm{d}r \gets \beta$.
    \State $s^2 \gets \frac{1}{n-2}\sum_k (\mathrm{CRPS}_k - \beta\bar{r}_k - \alpha)^2$, \quad $S_{rr} \gets \sum_k (\bar{r}_k - \bar{r})^2$.
    \State $\mathrm{se}(\beta) \gets \sqrt{s^2 / S_{rr}}$, \quad $\mathrm{se}(\alpha) \gets \sqrt{s^2\,(1/n + \bar{r}^2 / S_{rr})}$.
\State Report $\alpha$ as the estimated ceiling, and $\beta$ as a check on how strongly the estimate depends on $k$.
    \EndFor
    \Statex
    \Function{FairCRPS}{$y$, $S = \{s_1, \dots, s_m\}$} \Comment{fair $m(m-1)$ kernel}
    \State Sort $S$ into order statistics $s_{(1)} \le \cdots \le s_{(m)}$.
    \State \Return $\dfrac{1}{m}\displaystyle\sum_{i=1}^{m} \lvert y - s_{(i)} \rvert \;-\; \dfrac{1}{m(m-1)}\displaystyle\sum_{i=1}^{m} (2i - m - 1)\, s_{(i)}$
    \EndFunction
  \end{algorithmic}
\end{algorithm}

The second term of \textsc{FairCRPS} uses the same all-pairs identity as the exact floor in Equation~\ref{eq:crps_floor_sum}. Using the same all-pairs estimator for the ceiling, the floor and every model score is what makes the ratios in Table~\ref{tab:knn_slopes_limits} comparable. The only difference is that the floor divides by $N^2$ and not $N(N-1)$. This is a relative difference of $1/(N-1)$, or $5\times10^{-6}$ at $N = 200{,}000$ and about $10^{-4}$ in the smallest of the 14 $p_{\mathrm{T}}$ bins, which holds about 9{,}000 events. Mixing estimators would matter much more. If a model scored with the fair kernel were normalized by a floor estimated from an $m$-member marginal ensemble with the plug-in $m^2$ kernel, the reference would be inflated by $(m+1)/m$ \citep{ferro_effect_2008}. This adds $(1-\mathcal{S})/(m+1)$ of apparent skill, which is up to about $2\%$ at $m = 50$. That is about as large as the whole differential skill on momentum response and larger than it on $\Delta\eta$ and $\Delta\phi$.

The differential ceiling in Figure~\ref{fig:data_driven_limits} runs the same sweep on the query events of each of the 14 log-spaced parton-$p_{\mathrm{T}}$ bins, against the same per-flavour reference trees, and reads each result against the floor recomputed inside that bin, which removes the spectrum dependence of Section~\ref{sec:CRPS_framework}. Every sweep, global or per bin, uses the same neighbourhood ladder $\mathcal{K}$ on the first 200{,}000 test events of each split, with at most 400{,}000 reference events per flavour. Bins holding fewer than 200 test events are skipped. The whole procedure is one tree build per flavour plus $|\mathcal{K}|$ query passes per sweep, with no training and no gradients.

\end{document}